\documentclass[11pt,twoside]{article}
\usepackage{asp2006}
\usepackage{graphicx}
\usepackage{lscape}

\begin{document}
\title{Intrinsic Structure and Kinematics of the Sub-Parsec Scale Jet of M87}
\author{Y. Y. Kovalev}
\affil{
   Max-Planck-Institut f\"ur Radioastronomie,\\
                    Auf dem H\"ugel 69, 53121 Bonn, Germany; 
   \\
   Astro Space Center of Lebedev Physical Institute,\\
                    Profsoyuznaya 84/32, 117997 Moscow, Russia
}

\begin{abstract}
We report new results from a sub-parsec scale study of the inner jet in
M87 performed at 15 GHz with the Very Long Baseline Array. We have
detected a limb brightened structure of the jet along with a faint 3~mas
long counter-feature which we also find to be limb brightened.
Typical speeds of separate jet features are measured to be less than
0.05~speed of light, despite the highly asymmetric jet structure and the
implications of the canonical relativistic beaming scenario. The
observed intrinsic jet structure can be described in terms of a two
stream spine-sheath velocity gradient across the jet according to
theoretical predications based on the recently discovered strong and
variable TeV emission from M87. The jet to counter-jet flux density
ratio is measured to be greater than 200. The observed intrinsic jet
structure is broadly consistent with theoretical predictions of a
spine-sheath velocity gradient suggested by recently discovered TeV
emission from M87.
\end{abstract}

\section{Introduction} 
\label{s:intro}

The peculiar galaxy {M87} in the {Virgo} cluster was among the first to
be recognized as a powerful source of radio emission. M87 remains of
great interest today, since there is strong observational evidence for a
$3\times 10^9$ $M_{\sun}$ black hole located at the galactic nucleus
thought to power the relativistic jet \citep{H94, MMAC97}. Moreover, at
a distance of only 16~Mpc,  M87 is one of the  closest radio galaxies,
and as such it has one of the few jets which can be well-resolved on
sub-parsec scales in a direction transverse to the flow.

In this paper we report on observations made with the NRAO Very Long
Baseline Array (VLBA) at 2~cm wavelength. High dynamic range
images constructed from observations made in the year 2000 describe the
two-dimensional structure of the jet out to nearly 0.2~arcsec (16~pc)
and show the presence of a faint counter-feature. These observations
were complemented by regular observations made with lower sensitivity
between 1995 and 2007 to study the outward flow within the inner part of
the radio jet. 

\section{Jet Structure and Kinematics}
\label{s:obs}

M87 has been regularly observed with the VLBA since 1995 as part of the
2~cm VLBA survey \citep{2cmPaperI} and the more recent MOJAVE program
\citep{LH05}. In these programs, at each epoch we observed each source
for a total of about one hour, with multiple observations spaced over a wide 
range of hour angle. For M87, we have obtained a total of 23 images
between 1995 and 2007. Typically the rms noise in each image is about
0.3~mJy\,beam$^{-1}$.
We have supplemented these multiple epoch images using VLBA 2~cm archive
data from observations made at three epochs in 2000.
These later observations were made with full tracks in
hour angle each lasting about 10 hours using 2-bit recording at a
256~Mbps data rate.

\label{s:structure}

\begin{figure}[p]
\begin{center}
\resizebox{\hsize}{!}{
   \includegraphics[trim = 0cm 1cm 0cm 0cm]{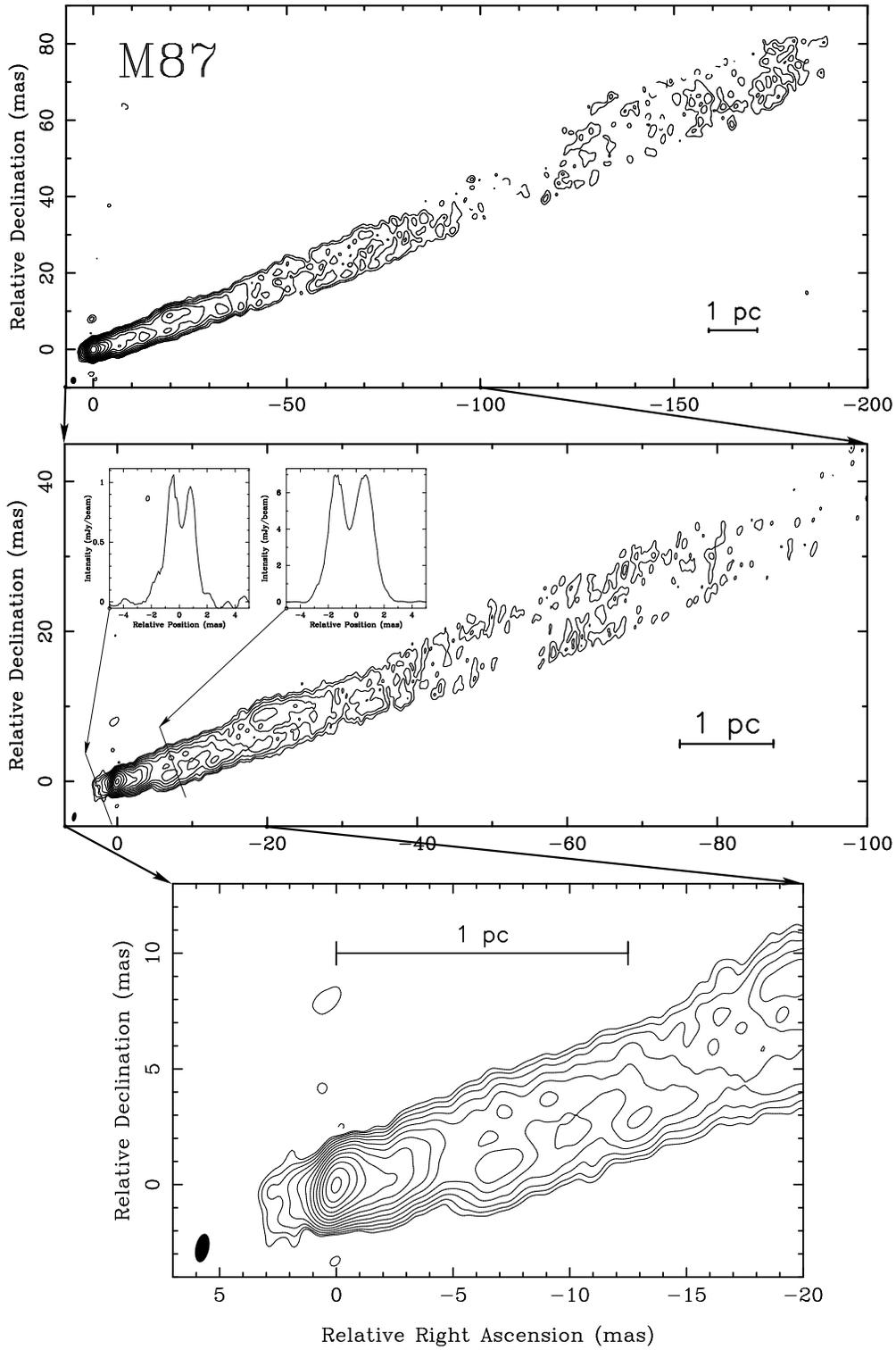}
}
\end{center}
\caption{\label{f:M87_VLBA}
VLBA 2~cm image of the M87 jet.
}
\end{figure}

In Figure~\ref{f:M87_VLBA}, we show the 2~cm image constructed from a
full track in hour angle made with the VLBA plus one VLA
element on 08 May, 2000. A tapered (upper) image shows structure out to
nearly 0.2~arcsec. The middle and lower plots present the naturally
weighted  CLEAN images of the inner jet. The beam is shown in the lower
left hand corner of each map. The contours are plotted in successive
powers of 2 times the lowest contour of 0.2 mJy/beam. The peak
intensity of the naturally weighted image is 1.00~Jy\,beam$^{-1}$, the
rms noise is 64~$\mu$Jy\,beam$^{-1}$,
and the corresponding dynamic range is
better than 15,000 to 1. The jet appears bifurcated, starting at about
5 mas (0.4~pc) from the core, characteristic of a single limb brightened
cylindrical or conical jet. The
M87 jet appears highly collimated, with re-collimation observed
between 2~pc where the opening angle is about $16^\circ$, and 12~pc where
the opening angle is only $6^\circ$ to $7^\circ$.
Figure~\ref{f:M87_VLBA} also shows the existence of weak 3~mas long structure
extending away from the bright core toward the southeast.
Indications of this counter feature were first suggested by 
\cite{Ly_etal04} on the basis of their 7\,mm VLBA observations.
This counter-feature also appears clearly bifurcated.
The two plots of the intensity profile cuts
of the counter-jet and the jet clearly show the limb brightening
(Figure~\ref{f:M87_VLBA}).

For several reasons we believe that the eastern extension may be the
counter-jet.  Based on their higher resolution 7~mm image, \cite{LWJ07}
have also detected the counter feature and have argued that the true
base of the jet cannot be offset by more than 2~mas from the bright
core; whereas we have detected the counter-feature to be at least
3.1~mas long.  Also, we note strong circular polarization at a
fractional level of $-0.5\%\pm0.1\%$ was detected by \cite{HL06}
coincident with the flux density peak, suggesting that this region of
the jet has an optical depth near unity \citep[e.g.,][]{J88} which is
characteristic of jet cores.  Since the eastern feature is more than 20
times fainter than the bright 1~Jy feature we have identified with the
core, it seems unlikely that it could be the actual core.  However,
considering that no other radio jet has been observed with comparable
linear resolution, sensitivity, and dynamic range, we cannot exclude the
possibility that we are seeing the detailed structure of the  optically
thick core of the jet, and that the counter-jet itself is not detected. 
Multi-frequency VLBI observations of the core region are
needed to determine if the eastern feature has a flat spectrum
characteristic of an optically thick synchrotron core or steep spectrum
as seen in other transparent jets.

\label{s:kinematics}

We find no evidence for motions faster than $0.05c$ within the inner 20
mas (1.6~pc). The fastest jet speed observed is only $0.024\pm0.004c$,
while the counter-jet apparently moves outward at $0.009\pm0.002c$.
Other features appear essentially stationary over the twelve years that
they have been observed, with nominal upper limits to their speed of
about $0.05c$. All features were detected at between 16 and 23 different
epochs. We believe that this interpretation is more robust than other
radio measurements which typically used only two or three epochs. Of
course, we cannot rule out the possible existence of a smooth
relativistic flow with no propagating shocks or other patterns.

\section{Discussion}
\label{s:discussion}

Considering that the Doppler boosted relativistic jet of M87 appears
one-sided out to kiloparsec scales, the absence of any clear motions in
the inner jet is somewhat surprising. We measure a jet to counter-jet
flux density ratio of between 10 and 15 (in the year 2000) at a position
between 0.5 and 3.1 mas from the core, and a lower limit of 200 between
3.1 to 6.0~mas away. Assuming that the jet is intrinsically
bidirectional and symmetric and that the bulk velocity flow is the same
as the pattern motion, our maximum reliable observed speed of
$\beta_\mathrm{app}=0.024\pm0.004$ would imply a jet to counter-jet
ratio near unity for the commonly accepted viewing angle of $\sim
40^\circ$ \citep[e.g.][]{OHC89, RBJM89}. If the jet and counter-jet are
intrinsically symmetric and oriented at an angle of $30^\circ$ to
$40^\circ$ to the line of sight, the observed jet to counter-jet flux
density ratios imply that the intrinsic flow speed, $\beta$, is 0.5 to
0.6 between 0 and 3.1~mas from the core and increases to $>0.9c$
beyond 3.1~mas. We conclude that the jet is either intrinsically
asymmetric, or there are no moving features within a rapidly flowing
plasma.

Apparent limb brightening is predicted from analytic and numerical
modeling of relativistic jets \citep{Aloy_etal2000,Perucho_etal07} and
can  be reproduced in terms of Kelvin-Helmholtz instability, which is
seen in extragalactic jets both at kiloparsec \citep[M87,][]{LHE03} and
parsec scales \citep[3C273,][]{LZ01}.  Limb brightening can be
particularly pronounced when the jet opening angle is greater than the
beaming angle \citep{GWD06} and especially if there is a velocity
gradient across the jet \citep{GDSW07}. A two layer ``spine-sheath''
model has been suggested to explain the existence of observed strong TeV
emission from BL~Lacs with apparently slow moving radio jets
\citep{CCCG00,PE04,G04}. \cite{SO02} and \cite{GTC05} have considered a
two component jet having a fast spine which produces the gamma ray
emission by inverse Compton scattering of the radio photons from a
surrounding slow layer or sheath. Spine-sheath models have also been
discussed by \cite{B96}, \cite{SBB98}, \cite{ARW99}, \cite{LB02}, and
\cite{Cohen_etal07}. The central gap seen in our VLBA image of M87, as
well as observed TeV emission \citep{A06} combined with the
lack of measurable motion within the inner 1.6~pc (\S\ref{s:kinematics})
appears to support this two component model. 

The recently reported detection of strong variable TeV emission from M87
\citep{A06} also presents a problem, since unless the radiating
plasma has a large Doppler factor, energy losses due to $\gamma -
\gamma$ pair production will extinguish the gamma ray emission
\citep[e.g.,][]{DG95}. However, we do not find any evidence for a fast
moving jet in M87 close to the central engine, but the observed TeV
emission can be explained in terms of a dual layer model with a fast
inner jet and a slower moving outer layer \citep{GTC05}. In this
picture, the inner jet is beamed away from us and is thus not seen in
our VLBA images, and we only observe the slower outer layer.  However,
even the slow outer layer must move at at~least $\beta>0.5$--$0.8c$ to
be consistent with the jet to counter-jet flux density ratio as
discussed earlier in this section.

\section{Summary}
\label{s:summary}

VLBA images of the inner radio jet of M87 show a limb brightened
structure  characteristic of a two stream spine-sheath velocity gradient
across the jet, as predicted from the recently discovered strong and
variable TeV emission and numerical modeling. Multi-epoch  VLBA
observations of seven separate jet features made since 1995 show typical
velocities  less than a few percent of the speed of light, contrary to
what might be expected from the highly asymmetric jet structure and the
canonical relativistic beaming scenario.  We have also found a weak
feature, located on the opposite side  of the bright core from the jet
which also appears limb brightened.  We believe this may be the
beginning of a counter-jet, but cannot exclude the  possibility that we
are seeing the detailed structure of the core.  We measure a jet to
counter-jet flux density ratio ranging from 10--15 close to the core to
more than 200 between 3 and 6 mas away.  Considering the lack of
observed motion in the inner jet, and the large jet to counter-jet
ratio, we conclude that either the jet of M87 is intrinsically
asymmetric or that the bulk plasma flow is much greater than any pattern
or shock propagation.

\acknowledgements

These proceedings are based on a paper by \cite{Kovalev_etal07}.
The author thanks M.~Lister, D.~Homan, and K.~Kellermann for their
collaboration.
The Very Long Baseline Array is a facility of the National Science
Foundation operated by the National Radio Astronomy Observatory under
cooperative agreement with Associated Universities, Inc. Part of this
work made use of archived VLBA and VLA data obtained by J.~Biretta,
F.~Owen, W.~Junor, and K.~Kellermann. YYK is a Research Fellow of the
Alexander von Humboldt Foundation. Part of this work was done by YYK
during Jansky postdoctoral fellowship at the National Radio Astronomy
Observatory. We thank D.~Harris, T.~Cheung, C.~Walker, M.~Perucho,
A.~Lobanov, E. Ros, and the MOJAVE team for helpful discussions and
contributions to this paper. This research has made use of NASA's
Astrophysics Data System.


\end{document}